\documentclass[10pt,a4paper]{article}

\def\heading #1{\bigbreak \begin{center} {\bf #1} \end{center}}

\title{Predicting atomic dopant solvation in helium clusters: the MgHe$_n$ case.}

\author{Massimo Mella\\
School of Chemistry, Cardiff University,\\
Main Building, Park Place, Cardiff CF10 3AB (UK), \\
Electronic mail: mellam@cf.ac.uk
\and Gabriele Calderoni\\
Dipartimento di Chimica Fisica ed Elettrochimica, \\Universit\`a degli Studi
di Milano, via Golgi 19, 20133 Milano, Italy\\
Electronic mail: gabriele.calderoni@unimi.it
\and Fausto Cargnoni \\
Istituto di Scienze e Tecnologie Molecolari (ISTM-CNR),\\
via Golgi 19, 20133 Milano, Italy\\
Electronic mail: fausto.cargnoni@istm.cnr.it}

\begin{document}

\bibliographystyle{prsty} 

\maketitle

\begin{abstract}
We present a quantum Monte Carlo study of the solvation and spectroscopic properties of the Mg doped
helium clusters MgHe$_n$ with $n=2-50$. Three high level (MP4, CCSD(T) and CCSDT)
MgHe interaction potentials have been used to study the sensitivity of the dopant location
on the shape of the pair interaction.
Despite the similar MgHe well depth,
the pair distribution functions obtained in the diffusion Monte Carlo simulations markedly differ for the three 
pair potentials, therefore indicating different solubility properties for Mg in He$_n$. Moreover, we found
interesting size effects for the behavior of the Mg impurity.

As a sensitive probe of the solvation properties, the Mg excitation 
spectra have been simulated for various cluster sizes and compared with the available experimental 
results. The interaction between the excited $^1$P Mg atom and the He moiety has 
been approximated using the Diatomics-in-Molecules method and the two excited $^1\Pi$ and $^1\Sigma$ MgHe
potentials.
The shape of the simulated MgHe$_{50}$ spectra show a substantial dependency on
the location of the Mg impurity, and hence on the MgHe pair interaction employed.

To unravel the dependency of the solvation behavior on the shape of
the computed potentials, exact Density Functional Theory has been adapted to the
case of doped He$_n$ and various energy distributions have been computed.
The results indicate the shape of the repulsive part of the MgHe potential as an important
cause of the different behaviours.
\end{abstract}

\pagebreak

\heading{I. INTRODUCTION}

The cold and gentle environment represented by bulk $^4$He and by $^4$He clusters has been attracting the attention
of the physical chemistry community due to very peculiar properties such as the absence of internal
friction, the small interaction energy with a doping impurity, and the ability to dissipate promptly the
excess energy of excited molecules (see Ref. \cite{tonnrev} for an extensive review on these subjects). 
These features
make the He nanodroplets an ideal environment to carry out reactions without the kinetic bottleneck
represented by the time needed by the reactants to diffuse and collide, and to record high accuracy
spectra of ultra-cold molecules and molecular complexes.

Despite the fact that the interaction energy between He atoms and the doping impurities is usually very
small, such interaction plays an important role in many interesting quantum phenomena. As an example we recall
the "adiabatic following" of the molecular rotations by the neighboring He atoms \cite{whaprl}. This 
effect induces the
increase of the momentum of inertia of molecules solvated by helium, and is experimentally detected by the
decrease of the spacing between the adsorption lines in the microwave spectrum \cite{tonnrev}.

Due to the highly quantum nature of He aggregates, even the much simple solvation process of a neutral
impurity is not completely rationalized, and up to now only few attempts to reach a detailed description of
the experimental findings have been carried out \cite{lehmann99,lehmann00}. Among the unsolved issues, the subtle interplay between
the various parameters of the system playing a role in the solvation mechanism (i.e. the features of the
interaction potential between helium and the impurity) still waits to be fully uncovered.

An attempt in this direction has been made by Ancillotto {\it et al.} \cite{anc}.
These authors modeled superfluid
helium by means of a Density Functional Theory (DFT) approach, using an approximate energy functional.
Assuming an  infinite atomic mass for the impurity, and that its interaction energy curve with helium takes
the form of a Lennard-Jones potential, they reached a clear cut description of the solvation phenomenon in
terms of the single dimensionless parameter $\lambda$, defined as:

\begin{equation}
\label{mgeq1}
\lambda = 2^{-1/6} \sigma^{-1} \rho \epsilon \mathrm{r}_e
\end{equation}

\noindent
where $\sigma$ is the surface tension of liquid He, $\rho$ is the number density of bulk He, $\epsilon$
and $r_e$ are the well depth and the equilibrium distance of the He-impurity potential, respectively. The
computed value of $\lambda$ unambiguously discriminates between opposite behaviours. Indeed, if $\lambda$ is
larger than 1.9 the free energy of the impurity decreases as it moves from the surface to the bulk helium,
indicating the onset of the solvation process. Conversely, if $\lambda<1.9$ the minimum free energy is
reached when the impurity resides on the surface of the bulk, and thus no solvation occurs.

Despite the merits of simplicity, and of reducing the number of independent variables to a single one, this
model was not devised to describe in detail the solvation process, but only to predict whether for a given
impurity the solvation occurs or not. Furthermore, the DFT approach does not take into account properly the
discrete nature and the anisotropic deformation \cite{ogata99,maxag} of the He aggregates, and this might lead to
overestimate the overall interaction energy with the impurity. The drawbacks of this approximation, as well
as the ones cited above, are expected to be particularly relevant in systems with $\lambda$ close to the 
critical value 1.9.

A deep description of the solvation process can be gained by solving
exactly the Schr\"odinger equation of nuclear motion using an explicit many-body
algorithm. In systems where accurate interaction potentials between helium and the impurity are available,
the quantum Monte Carlo (QMC) approach is probably the best suited, and has been already applied
successfully to the study of doped helium clusters 
\cite{ogata99,maxag,yamashita,maxheh-,max3he,gianturcoheh-,smallOCS}. Though the
DMC simulations cannot recover the temporal evolution of the system,  they provide many important quantities
that are hardly accessible experimentally, such as radial and angular distribution functions, solvation
energies, excitation spectra, as well as their dependence on the size of the helium aggregate.

In the present study we applied this method to study the doping of helium clusters with a neutral Mg atom, a
system that in recent years has been the subject of two experimental investigations \cite{morika99,reho00}. 
Moriwaki {\it et al.} \cite{morika99} measured the emission and the excitation spectra of Mg dispersed 
in liquid helium by means of
ultraviolet laser excitation, and focused their attention on the signals arising from the
Mg(3s$^2)^1$S$\rightarrow$Mg(3s3p)$^1$P transition. Reho {\it et al.} \cite{reho00} studied 
this same transition of Mg while interacting with
helium nanodroplets. Based on the comparison with the available experimental data for Mg  \cite{morika99} and for other
neutral metal atoms in helium, Reho {\it et al.} concluded that Mg is indeed solvated by helium, and thus does not
reside on the surface of the nanodroplets. However, they stressed that no clear cut description of this
system could be recovered by theory using the model proposed by Ancillotto {\it et al.} \cite{anc}
because the $\lambda$
values computed from the available model potentials for Mg-He are heavily scattered around the critical value
of 1.9 (see Refs. \cite{stienk_casr} and \cite{partridge01}, and Ref. \cite{reho00} for an exhaustive review up 
to 1999); furthermore, the performance of
this scheme in nearly critical conditions (i.e. for $\lambda$ values close to 1.9) still needed to be
assessed.

To shed light on these arguments is the main scope of the present work. 
As a first step we generated accurate potential energy surfaces (PES) for the interaction between a Mg atom in 
the ground state and He using the MP4 and CCSD(T) methods, and for Mg in the $^1$P state and He 
with a multiconfigurational approach, adopting for all computations high quality basis sets. The 
PES of the excited Mg-He complex are necessary to compute the excitation spectrum of Mg attached to He 
clusters. While our investigation was on its way, two alternative potentials for the ground state Mg-He 
interaction were proposed, based on CCSD(T) \cite{partridge01} and CCSDT \cite{hinde03}
computations. The interaction energies computed by Partridge {\it et al.} \cite{partridge01} were nearly 
superimposable to our CCSD(T) potential, while the ones by Hinde \cite{hinde03} turned out to 
be noticeably different from both potentials computed by us; therefore we 
extended our investigation by adopting also Hinde's proposal. 

Second, we detailed the solvation 
phenomenon by performing DMC simulations on MgHe$_n$ clusters with growing size (n=2-50);
this approach helps in highlighting possible size effects that are difficult to probe experimentally.
Since the manifolds of the excitation spectrum largely depend on
the onset of solvation, we also computed the PES for the three low-lying
Mg(3s3p)-He excited states. Their implementation in DMC simulations according to the Diatomics-in-Molecules
(DIM) \cite{ellison63} scheme allowed to recover the excitation spectrum of magnesium interacting with helium clusters,
and thus to compare directly theoretical with available experimental results.

The outline of this work follows. Section II presents the details of the ab-initio computations performed to
determine the ground and excited states two-body Mg-He interaction potentials, with a short discussion on the
relevance of the three-body effects in MgHe$_n$ clusters. In Section III we give a short introduction to the
quantum Monte Carlo methods used in this work to solve the Schr\"odinger equation for nuclear motion. In
Section IV we present the results of the DMC simulations, and compare our data with the available experimental 
measurements \cite{morika99,reho00}. Finally, Section V reports our conclusions, along with a prospect on
future applications.

\heading{II. INTERACTION POTENTIALS}

The availability of accurate interaction potentials is a prerequisite for a reliable modeling of 
doped helium clusters. In the present investigation, the complete PES 
of the ground and excited states Mg-He$_n$ clusters are approximated by means of two-body 
terms. Within this approach the interaction energy in the ground state clusters is 
predicted simply by summing up all the pairwise Mg($^1$S)-He and He-He contributions. Conversely, 
in the excited clusters we adopted the DIM formalism \cite{ellison63} to recover 
the overall Mg-He$_n$ interaction energy from the computed two-body Mg($^1$P)-He potentials.

As concerns the He-He interaction, we selected the TTY potential proposed by Tang et al.
\cite{tang95}, which is established as one of the most accurate PES for this system. 

In the case of the ground state Mg-He complex, a number of interaction energy curves were
proposed in the last fifteen years \cite{reho00,partridge01,hinde03,czuchaj87,funk89,hui97,kleine00}.
The most accurate were computed by Funk {\it et al.} \cite{funk89}, Partridge {\it et al.}
\cite{partridge01}, and by Hinde \cite{hinde03}. Funk {\it et al.} performed MP4 computations 
adopting a basis set derived from Huzinaga {\it et al.} \cite{huzinaga85} and augmented with diffuse 
functions; these authors provided a well depth estimate of 4.54 cm$^{-1}$ at the Mg-He internuclear 
distance of 5.16 {\AA}. Partridge {\it et al.} carried out CCSD(T) computations with the aug-cc-pVQZ 
basis set for both He and Mg \cite{dunning94}, supplemented with the 332 set of bond functions proposed 
by Tao and Pan \cite{tao92}. They obtained a well depth of 4.76 cm$^{-1}$ at the distance of 5.09 {\AA}. 
More recently, Hinde combined valence-only CCSDT computations of nearly Full Configuration Interaction (FCI)
quality with a core-valence correction estimated at CCSD(T) level and, using high quality basis 
sets, he proposed a minimum interaction energy of 5.01 cm$^{-1}$ at the internuclear distance of 
5.07 {\AA}. In the present investigation, we adopted the potential given by Hinde, and we also computed 
two new MP4 and CCSD(T) PES adopting a larger basis set \cite{funk89} and a finer spatial 
grid \cite{funk89,partridge01} than the available investigations with these same 
theoretical schemes. 

Contrary to the ground state complex, the interaction between helium and 
magnesium in the Mg(3s3p) excited state has received much less attention. Indeed, 
to the best of our knowledge the only data available is 42 cm$^{-1}$ for the well 
depth of the $^1$P state, but this single value was reported by Moriwaki et al. \cite{morika99} 
without any details on the method of calculation. We therefore determined the 
complete potential energy curves of the three Mg(3s3p)-He states by performing MR-CI 
computations, since these potentials are necessary for the implementation of the DIM 
scheme \cite{ellison63}. 

All the {\it ab initio} computations were carried out with the GAMESS \cite{gamess} 
and the Gaussian \cite{g98} suite of programs; all interaction energy data were
corrected for BSSE with the counterpoise scheme proposed by Boys and Bernardi \cite{bbcp}.

\heading{a. Choice of the basis set.}

To define an appropriate set of basis functions, we performed a series of test MP4 computations 
in the region of the PES minimum, keeping frozen only the 1s electrons of Mg. For the He atom, we adopted 
the d-aug-cc-pVnZ (n=3-5) sets proposed by Woon and Dunning \cite{dunning94}; for Mg we 
considered three alternatives: the 6-311+G3df \cite{mclean80}, the Roos augmented 
triple-zeta A.N.O. \cite{widmark91}, and the basis set proposed by Archibong and Takkar \cite{archi91}. 
We also tested the effect of including bond functions, placed at midway between helium 
and magnesium, using alternatively the 332 set of Tao and Pan \cite{tao92} and the 33221 set by 
Cybulski and Toczylowski \cite{cybulski99}. 

The results for the combinations of the basis set considered are reported in Table \ref{tab1} 
and can be summarized as follows: i) whatever the basis set adopted for Mg and He atoms, 
the inclusion of bond functions lowers significantly the well depth; ii) given a set for 
the two atoms, the well depths obtained with the 332 and the 33221 bond functions agree 
within 0.05 cm$^{-1}$; iii) once a basis set for He is chosen, and a set of bond 
functions is included, the differences among the three sets for the Mg atom are limited 
to 0.2 cm$^{-1}$; iv) to use the d-aug-cc-pVQZ or the d-aug-cc-pV5Z set for He is nearly 
equivalent, provided that a set of bond functions is included. 
We also tested that these same conclusions holds true at the CCSD(T) level of theory.
We finally devised to use the 6-311+G3df basis set for Mg, the d-aug-cc-pVQZ set for He, 
and the 332 set of bond functions, this choice representing the smallest basis set that 
provides interaction energies nearly converged.

\heading{b. Ground state PES}

The ground state PES for Mg-He has been computed at the MP4 and CCSD(T) levels of theory, keeping 
frozen only the 1s electrons of magnesium. We considered 25 internuclear distances, ranging 
from 4.0 to 10.0 {\AA} with 0.25 {\AA} steps. Within this interval the energy data have been 
interpolated by means of natural quintic splines, whereas the long range potential has been 
expressed with the analytical form $-C_6/r^6$. 

The computed energy curves for Mg-He are shown 
in Figure 1 along with the one proposed by Hinde \cite{hinde03} and the He-He TTY potential. 
The MP4 and the CCSD(T) PES exhibit a minimum interaction energy of -5.70 and -4.75 cm$^{-1}$, respectively, 
at the internuclear distances of 5.03 and 5.11 {\AA}. Our CCSD(T) potential is nearly superimposable to 
the one presented by Partridge {\it et al.} \cite{partridge01} using the same level of theory, while 
our MP4 estimate is significantly more attractive as compared to the one determined by 
Funk {\it et al.} \cite{funk89}. As for the potential proposed by Hinde \cite{hinde03}, it is 
intermediate between CCSD(T) and MP4 results ($\epsilon$=5.01 cm$^{-1}$, r$_e$=5.07 {\AA}), being much closer to 
the former. Finally, it is noteworthy that the computed MP4 potential is about 1 cm$^{-1}$ deeper than 
the CCSD(T) one computed with the same basis set, contrary to the common experience for van der Waals systems. 
Funk {\it et al.} \cite{funk89} suggested that this feature is due to the quasi-degenerate effects 
proper of the alkaline-earth metal atoms.

The gross features of these potentials allow to draw a first picture of the behaviour of Mg in He 
using the scheme proposed by Ancilotto {\it et al.}  \cite{anc}. At variance with the uncertain 
situation encountered by Reho {\it et al.} \cite{reho00}, all the new generation PES provide 
the same clear cut prediction. Indeed, the $\lambda$ values range from 2.66 (CCSD(T) potential) 
to 3.14 (MP4), and a value of 2.78 is found for the potential by Hinde. These data are well beyond 
the critical value of 1.9, and even considering the correction proposed to account for the zero point 
motion energy of the impurity \cite{anc}, we may assert that the model unambiguously foresees the Mg 
atom to be solvated by helium.

\heading{c. Excited states PES.}

The first singlet excited state of a free Mg atom is (3s3p) $^1$P, and its threefold degeneracy 
is removed upon interaction with a He atom, thus splitting in two degenerate $^1\Pi$ and a 
single $^1\Sigma$ state. Roughly speaking, in the $^1\Pi$ states one electron of magnesium 
lies in a p orbital orthogonal to the Mg-He internuclear axis, and consequently He essentially 
interacts with a positive ionic core: we expect the PES to exhibit a deep well at short 
distances. Conversely, in the $^1\Sigma$ state the excited electron lies in the p orbital 
pointing toward helium, thus generating a potential largely repulsive even at long Mg-He 
distances. The PES of these states were determined by means of CAS (4 electrons in 10 orbitals) 
computations, followed by a multi-reference single and double substitutions configurations 
interaction. The CAS scheme included in the active space the 3s, 3p and 4s orbitals of Mg, and 
the 1s, 2s, 2p orbitals of He. These computations were density averaged over the first four 
states. 

The MRCI method is not size consistent, and therefore the interaction energies have been 
calculated as $E=E_{\mathrm{MgHe}}-E_{\mathrm{Mg}}-E_{\mathrm{He}}+\mathrm{ESC}$, where 
the last term is a size consistency correction defined as $\mathrm{ESC}=E_{\mathrm{MgHe}}(r=\infty)-
E_{\mathrm{Mg}}-E_{\mathrm{He}}$. To check for the soundness of this approach, we carried 
out an extensive comparison between MRCI and FCI computations, considering two combinations 
of basis sets: 6-31+G*/aug-cc-pVTZ and 6-311+G3df/d-aug-cc-pVTZ, where A/B indicates the set 
for Mg and He, respectively. As it can be seen from Figures 2 and 3, the overall agreement 
between MRCI and FCI results is very satisfactory for the two basis set choices. We are therefore 
confident that even the use of a more extended basis set would not alter this agreement, and 
thus we determined the three PES at MRCI level using the same basis set as for the ground 
state case. The interaction energies have been determined on 26 Mg-He arrangements, ranging 
from 3.0 to 12.0 {\AA}; the analytical representations are obtained according to the same 
strategy adopted for the ground state. The $^1\Pi$ potential reported in Figure 2
exhibits a well depth of 39.58 cm$^{-1}$ at the equilibrium distance of 3.82 {\AA}, whereas the 
$^1\Sigma$ curve is repulsive from 7.79 {\AA} inward, and has a very shallow well of 0.81 cm$^{-1}$ 
at 8.62 {\AA}, as reported in Figure 3. Finally, the asymptotic separation between the excited 
levels and the ground state was set to the experimental value of 35051.264 cm$^{-1}$ \cite{jphyschem}.

\heading{d. Relevance of the many-body contributions.}

As mentioned in the introduction, in the DMC simulations the ground state PES of MgHe$_n$
has been approximated as a sum of pairwise interactions. Among the contributions excluded, the 
major role is certainly played by the three-body effects. In systems scarcely polarizable 
such as the helium aggregates, the most important three-body component arises from the non 
additivity of both the exchange and dispersion terms, and the former is generally much larger 
than the latter. Exchange contributions are typically attractive in triangular arrangements and 
repulsive in collinear geometries, while the dispersion ones behave oppositely. 
In helium based systems these effects are usually quite small, and therefore the common practice 
is to neglect them regardless of the doping impurity.

Nevertheless, we carried out a series of test computations to check the validity of this 
approximation in Mg doped He clusters. We considered MgHe$_2$ complexes with C2v symmetry, since 
the highest three-body contributions arise in triangular arrangements. All calculations 
have been performed at the MP4 level of theory, using the same basis set choice as for 
the two-body potentials but for the exclusion of the bond functions, that proved nearly negligible 
in this respect. In Figure 4 we report the two- and three-body contributions to the interaction energy 
as scans along the HeMgHe angle at fixed Mg-He internuclear distances. 
The three-body terms are not negligible only at short Mg-He distances and small He-Mg-He angles.
In these arrangements the two body PES is only slightly attractive or even repulsive, and therefore
we are confident that the inclusion of many-body effects in the MgHe$_n$ complexes would
scarcely affect the results of the DMC simulations.

\heading{III. METHODS}

It is well known that pure and doped He clusters are characterized by 
a highly quantum nature, a feature manifesting itself in a
small total binding energy and a wide anharmonic motion of both the 
doping impurity and the He atoms. As a consequence of the intrinsic anharmonicity
and of the experimental size of these clusters (usually of the order of several thousands atoms),
the possibility of using either the harmonic approximation or more accurate basis set/ grid-bases 
approaches is usually hindered.

To describe at atomistic level the solvation properties of doped clusters and to
compute their excitation spectra, we 
believe that the quantum Monte Carlo (QMC) methods are the best suited techniques. 
Since these methods are well described in the literature \cite{reybook}, we 
restrain ourselves from presenting long discussions, except for the technical 
details that are relevant to the present work. 
Here, we employed variational Monte Carlo (VMC) to 
optimize a trial wave function $\Psi_T(\mathbf{R})$
and diffusion Monte Carlo (DMC) to correct the 
remaining deficiencies of $\Psi_T(\mathbf{R})$, projecting out all the 
excited state components and sampling $f(\mathbf{R}) = \Phi_0(\mathbf{R})\Psi_T(\mathbf{R})$.
In both cases a description of the ground state is sought, the low temperature of the
clusters (0.37 K) and the large energy gap between vibrational excited states in He$_n$ suggesting
that thermal excitations should not play a relevant role (for a discussion on this topic see
Ref. \cite{maxag}).

In atomic units, the Hamiltonian operator for MgHe$_n$ reads as
\begin{equation}
\label{mgeq2}
{\mathcal H} = -\frac{1}{2} \left( \sum _{i=1}^{n}
\frac{\nabla _i ^{2}}{m_{^4 \mathrm{He}}} + \frac{\nabla_{\mathrm{Mg}}^{2}}
{m_{\mathrm{Mg}}} \right) + V(\mathbf{R})
\end{equation}
\noindent
As mentioned before, we assume a pair potential of the form
$V(\mathbf{R}) = \sum _{i<j} V_{\mathrm{HeHe}}(r_{ij}) + \sum _i
V_{\mathrm{MgHe}}(r_{i\mathrm{Mg}})$ for the clusters with the magnesium atom in the $^1$S
electronic ground state.

Our trial wave function has the common form
\begin{equation}
\label{mgeq3}
\Psi_T(\mathbf{R})=\prod _{i<j}^N\psi(r_{ij})\prod_i^N\phi (r_{i\mathrm{Mg}})
\end{equation}
where no one-body part was used, and \cite{Rick}
\begin{equation}
\label{mgeq4}
\psi (r) = \phi (r) = \exp [ -\frac{p_5}{r^5}
-\frac{p_2}{r^2} -p_1 r -p_0 \ln(r)]
\end{equation}

The parameters of the model wave function were fully
optimized minimizing the mean absolute error of the local energy $E_{loc}(\mathbf{R}) =
\mathcal{H}\Psi_T(\mathbf{R})/\Psi_T(\mathbf{R}) ={\mathcal H}_{loc}(\mathbf{R})$ over 
a fixed set of points as proposed in Ref. \cite{robust}.  The optimized
wave functions were successively employed to guide the DMC simulations of
the doped clusters in order to sample the mixed distribution
$f(\mathbf{R})=\Phi_0(\mathbf{R})\Psi_T(\mathbf{R})$.
These distributions were used to compute exactly the energy values using the
mixed estimator
\begin{equation}
\label{mgeq5}
\langle {\mathcal H} \rangle_M = \frac{\int f ({\mathbf R})
{\mathcal H}_{loc} ({\mathbf R}) d{\mathbf R}}
{\int f ({\mathbf R}) d{\mathbf R}}
\end{equation}
\noindent
as well as the mixed and second order estimate (SOE) $\langle{\mathcal O}\rangle_{SOE}
=2\langle{\mathcal O}\rangle_{M}-\langle{\mathcal O}\rangle_{VMC}$ of many other
expectation values (e.g. the interparticle distribution functions).
The SOE was used to reduce the bias introduced in the mixed 
estimate of operators that do not commute with the Hamiltonian by the 
use of a non-exact trial wave function. 

As for the absorption spectrum of the Mg atom, we computed it
with the same semiclassical approach used in Ref. \cite{maxag} to compute the
Ag spectrum, adapting a technique previously proposed by Cheng and Whaley \cite{cheng} for the 
Franck-Condon line shapes of an electronic transition in a condensed phase system. 
The method was originally presented by Lax \cite{lax} and modified to take into 
account the system temperature of 0 K. 
In its crudest approximation, the spectral lines of a chromophore are computed 
collecting the distribution of the differences $V_{exc}(\mathbf{R})-V_{gs}(\mathbf{R})$
over the sampled $f(\mathbf{R})$. 
In our case, $V_{gs}(\mathbf{R})$ ($V_{exc}(\mathbf{R})$) is the interaction 
potential between the ground (excited) state Mg atom with the He atoms. 
The three electronic states for the excited Mg attached to the He cluster are obtained from the two
dimer excited potentials $^1 \Sigma$ and $^1 \Pi$ using the
DIM method \cite{ellison63}. Since all the details needed to implement it are well described
by Nakayama and Yamashita \cite{yamashita}, we refer to their paper and to Ref. \cite{maxag}
for further discussions, especially related to the general accuracy of the method.

\heading{IV. RESULTS}

\heading{a. Energetics}
Due to the highly quantum nature of He$_n$, we limit our presentation of the MgHe$_n$ energetics
only to the fully quantal DMC results and postpone a short discussion on the potential energy values
of the global minimum structures to the next section.

The DMC energies $E(n)$ and the differential values $\Delta(n)=[E(n)-E(m)]/(n-m)$ obtained using 
the three potentials for the MgHe$_n$ clusters are presented in Table \ref{tab2}.
Here, MgHe$_m$ is the largest cluster for which $n>m$ still holds.
From these energetic results, it appears that the CCSD(T) and the CCSDT data
give a very similar description of the clusters, while the MP4 produces slightly 
lower total energies as expected on the basis of the deeper well.

The $\Delta(n)$ values can be interpreted as the negative of the He evaporation energies and 
are shown in Figure 5.  From this, we notice that for all three interaction potentials $\Delta(n)$
monotonically decrease upon increasing $n$, a finding that is different from what is usually found
for impurities interacting strongly with He \cite{maxag,largeOCS,largeHF},
and for which a multiple-shell effect is usually found. Instead, the behavior of $\Delta(n)$ for MgHe$_n$
is quite similar to the one obtained for pure He$_n$ \cite{shelleffect} (also shown in Fig. 5)
and for dopants floating on the surface \cite{maxheh-,gianturcoheh-}.
In particular, our CCSD(T) and CCSDT $\Delta(n)$ values for the large clusters are very close to
the pure cluster results presented in Ref. \cite{shelleffect}.
The MP4 potential gives somehow larger values for the same quantity, a difference in touch with 
the slightly larger interaction energy computed by the MP4 method and that
might induce a more compact structure for the He moiety.

Another interesting quantity is represented by the binding energy of Mg to the helium cluster
$\mathrm{BE}(n)= E^{\mathrm{He_n}}-E^{\mathrm{MgHe_n}}$ (often indicated as the dopant
chemical potential) shown in Table \ref{tab3} for He$_n$ up to $n=40$. 
These values were obtained using the DMC energies
for the pure helium clusters presented in Ref. \cite{maxag}.
The energy data obtained with the MP4 potential monotonically increase with cluster size, and appear to be
nearly converged at n=40. A similar behaviour is found using the CCSDT and CCSD(T) potentials up to n=30, but
in these cases there is an unexpected decrease of BE when the largest clusters (n=40) are considered.
This different behaviour of BE is due to relevant differences in the structures of the doped clusters
depending on the potential adopted and the cluster size, as
will be discussed in the next section.

\heading{b. Structure}
Let us start considering the relative shape of the MgHe and HeHe interaction curves shown in Figure 1. It
is clearly evident that the He-He curve has a deeper minimum $\epsilon$ and a shorter equilibrium distance
$r_e$ than all the Mg-He ones. Similar situations have been previously investigated \cite{cozzini} and
it is easy to predict that the largest species should segregate on the surface to 
reduce the strain in the optimized structure.
It is therefore quite likely that a classical minimization of the 
cluster interaction potential would produce global minimum structures 
showing a compact He moiety with Mg
lying outside this core, as already found for Ne$_n$H$^-$ \cite{gianturconeh-}
and He$_n$H$^-$ \cite{gianturcoheh-}. To confirm this tendency for our systems, we minimized 
the interaction potential starting from several
thousands random configurations for each cluster, invariably obtaining a "floating" Mg
impurity and a compact He moiety as lowest energy structure. As a consequence, the solubility of Mg in He$_n$ 
suggested by the cluster experiments \cite{reho00} must be considered a purely 
quantum phenomenon.

To extract the structural properties of the Mg doped clusters from DMC simulations, we computed several 
average values and distribution functions for the particle-particle distances and for the distance between
a particle A (either He or Mg) and the (geometrical) centre of the He moiety (gc)

\begin{eqnarray}
R_{\mathrm{gcMg}}=\| \mathbf{r}_{\mathrm{Mg}}-\sum_{i=1}^n \mathbf{r}_i/n\| \\
R_{\mathrm{gcHe}}=\| \sum_j^n (\mathbf{r}_j-n^{-1}\sum_{i=1}^n \mathbf{r}_i)/n\| 
\label{mgeq6}
\end{eqnarray}

Figure 6 shows the behavior of the average $R_{\mathrm{gcMg}}$ as a function of the number of 
He atoms in the cluster for the three interaction potentials used in this work. As for the MP4 potential,
$\langle R_{\mathrm{gcMg}} \rangle$ shows a steep decrease
upon increasing $n$, strongly resembling the case of AgHe$_n$ \cite{maxag} where this behavior
indicated the onset of solvation of the Ag atom in the He clusters.
Conversely, the two CC-based pair interactions produced somehow unexpected and peculiar behaviors that,
to the best of our knowledge, have never been found before with other dopants. More precisely, both
the CCSD(T) and CCSDT potentials generated almost constant $\langle R_{\mathrm{gcMg}} \rangle$ values
for $n\leq 30$; this similar behavior is then followed by a sudden increase for CCSD(T) and a less
rapid decrease in the case of CCSDT. Indeed, the trends shown by both CCSD(T) and CCSDT are somewhat
intermediate between the one exhibited by the impurities undergoing solvation (with $R_{\mathrm{gcMg}}$
decreasing as a function of $n$) and the ones floating on the helium droplet surface (in these systems
$R_{\mathrm{gcMg}}$ monotonically increases with increasing $n$) as in the case of
H$^-$He$_n$ \cite{maxheh-,gianturcoheh-}.
These unusual features are hardly interpreted on the basis of the available literature results, and to
put them in relation with a definite cluster structure - particularly concerning the onset 
of solvation - requires a more detailed treatment of the DMC data.

Figure 7 shows the pair distribution function $p(R_{\mathrm{gcMg}})$ for several MgHe$_n$
clusters and for any of the PES adopted: MP4, CCSD(T) and CCSDT.
The curves in the figure are normalized so that $4\pi\int_0^{\infty} r^2 p(r) dr = 1$. 
As clearly seen in Fig. 7a, the simulations with the MP4
potential produced a solvated dopant sitting in the centre of the large He clusters, the smaller ones
showing a slightly different behavior due to the incomplete first solvation shell of Mg. The simulations 
carried out using the CCSD(T) potential (Fig 7b) gave instead a largely different prediction of the solvation
properties: for $n\leq30$, the distributions present a peak with the maximum in the range of 7-8 bohr, therefore 
locating Mg relatively far away from the centre of the droplet and explaining the almost constant
$R_{\mathrm{gcMg}}$ values; for $n=40$ and 50, the peak maximum
is displaced further away from the He moiety centre, clearly indicating the lack of solvation for Mg. 
As for the CCSDT results (Fig 7c), the behavior of $p(R_{\mathrm{gcMg}})$ is even more complicated.
For $n\leq20$, the distributions show a single peak whose maximum is located in the range of 7-8 bohr similarly
to the CCSD(T) ones. Upon increasing $n$, they first assume a bimodal shape ($n=25$ and 30), and then
transform their shape building a single broad peak centered at $R_{\mathrm{gcMg}}=0$.



As mentioned before, the dependency of the solvation properties on the number of He atoms shown by both 
the CCSD(T) and CCSDT Mg-He pair interactions has never been observed for any previously studied dopant.
In the CCSDT case, our findings suggest the onset of a dynamical many-body effect stabilizing
the solvated dopant more than the surface one after a critical number of heliums is reached. 
A likely explanation for this behavior
is obtained considering what happens displacing a Mg atom from the surface to the interior of the droplet.
During this process, a cavity is formed inside the cluster due to the incoming Mg, this cavity and the
displaced He atoms contributing to increase the droplet external surface area by a quantity $\Delta S$.
Upon increasing $n$ (and hence the total volume of the He moiety), the absolute value of 
$\Delta S$ is reduced generating, as a consequence, a smaller increase in "surface energy" due 
to the lost He-He interactions, and making more favorable the dopant solvation \cite{surface_volume}.
It must be stressed, however, that the dopant solubility depends also on the shape (e.g. well depth and location) 
of the interaction energy between He and Mg: if the energy loss due to the formation of the cavity and the increase in
surface area is not overcompensated by the stronger interaction between the solvated dopant and the droplet, the
impurity will be segregated on the surface as a way to reduce the total energy of the system.
So, the cause for the different behavior of the CCSD(T) and CCSDT distributions must be sought in the slight 
differences presented by the two interaction potentials.

A somehow different perspective of the structural features of MgHe$_n$ 
is given by the distributions of the cosine values for the HeMgHe angle presented
in Figure 8 where, again, different behaviors are shown. In the case of the MP4 interaction (Fig. 8a), the 
distributions show a gradual change upon increasing the number of He atoms: the pronounced features present 
in the distribution for the small clusters ( i.e. the sharp maximum around $\cos(\mathrm{HeMgHe})=0.8$ 
and the two minima at $\cos(\mathrm{HeMgHe})=$ 1 and -1) smooth for larger $n$ producing, in the case of MgHe$_{50}$,
a substantial raise of the minimum at $\cos(\mathrm{HeMgHe})=1$, a displaced short maximum around 
$\cos(\mathrm{HeMgHe})=0.85$, and a plateau at lower values of the cosine. A similar behavior was
exhibited by the AgHe$_n$ clusters, an evidence of the complete solvation
of the dopant and of the formation of a compact and ordered first shell of He atoms around the silver atoms.
Furthermore, the raise of the minimum at $\cos(\mathrm{HeAgHe})=1$ indicated the
onset of the second solvation shell for Ag. The same process occurs also for MgHe$_n$ when the MP4 data are considered,
though a substantial difference between Mg and Ag doped clusters occurs. Indeed, in AgHe$_n$ the plateau
at low cosine values is present even in small clusters (e.g. AgHe$_6$), while in MgHe$_n$ it becomes
evident only for clusters with at least 25 He atoms. This finding is probably a consequence of the smaller
$\epsilon$ and the larger r$_e$ in the
MgHe interaction potential than in the AgHe one, the shallower interaction curve allowing 
the He atoms in MgHe$_n$ to be less tightly bound to the dopant and to
cluster together in more compact way. In turn, this retards the formation of an angularly uniform 
first solvation shell around the dopant.

Figure 8b shows the cosine distribution functions for the CCSD(T) curve, two striking differences
being apparent with respect to the previous results. First, only small changes in the distributions are produced
upon increasing the number of He atoms; this indicates the presence of a similar structure for 
the all the clusters.
Second, the density at $\cos(\mathrm{HeMgHe})=1$ is substantially different from zero 
already at $n=8$ and increases for larger $n$; this feature suggests that two He atoms can be 
found along the same radius departing from Mg even in the small clusters, and
these angular distributions would be consistent with a dopant floating on 
the surface of the He droplet whose radius increases upon increasing the number of He atoms.
At variance with this interpretation, Figure 8b also shows a substantial density at 
$\cos(\mathrm{HeMgHe})=-1$ for the small clusters suggesting the presence of a less "clear-cut" 
structural situation that can be explained either by imaging
the Mg atom as "sitting" in a deep dimple on the surface and having a fluctuating "crown" 
of He atoms in equatorial position or, alternatively, by
invoking large excursions of one helium away from the cluster surface and around the dopant.

Figure 8c shows the probability densities obtained during the DMC simulations employing the CCSDT pair interaction.
For $n\leq 30$, these distributions show an almost identical shape and are quite similar to the ones obtained using 
the CCSD(T) curve, therefore indicating the presence of a strongly anisotropic environment around the Mg atom. 
Upon increasing $n$, the distributions change smoothing the maximum around $\cos(\mathrm{HeMgHe})=0.9$
and increasing the density at lower values of $\cos(\mathrm{HeMgHe})$, a finding which is in phase with
the behavior of the radial probability density previously described. However, the absence of the distribution plateau 
characterizing the MP4 results at low cosine indicates that the He environment around Mg is still far from 
being isotropic.

All the structural information discussed so far can, somehow, be condensed in the more pictorial 
representation provided by snapshots of the cluster geometries obtained during the DMC simulations with the 
three pair interactions; Figure 9 shows typical configurations sampled during the simulations of the 
largest clusters. Whereas the configuration extracted from the MP4 simulation (Fig. 9a) clearly shows a Mg completely
surrounded by He atoms, walkers from CCSD(T) (Fig. 9b) show Mg preferentially
resident in a quite profound dimple.
A less clear-cut situation was found for CCSDT data: although a completely solvated, but slightly off-centered, 
dopant is the most likely geometry extracted from the DMC simulations (Fig. 9c), we also found clear evidences 
that Mg can and does reach the surface of the droplet during the simulations. 

The overall picture given by the results discussed in this section is very intriguing. First, we found that
even very small differences among the potentials adopted in the DMC simulations produce quite different
structures in the doped clusters. Second, the onset of the solvation itself is highly sensitive to very small
changes in the interaction potential of the dopant and to the number of He atoms contained in the droplet. In 
borderline systems it is therefore quite restrictive to
describe the solvation as simply occurring or not, and the properties of the dopant are very hard to classify
(see the DMC simulations on MgHe$_n$ adopting the CCSD(T) and CCSDT PES). Further discussion on the role of the
Mg-He interaction energy curve in several components of the solvation phenomenon is outlined in Section IV.
\nocite{Taken all together, the results obtained using the three different potentials are quite surprising, especially
considering that only small differences are present between the various interaction curves. In turn, this
finding also suggest that quantum solvation in He clusters is highly sensitive to small changes
in the total potential for atomic dopants, expecially when they represent borderline cases.
Further discussions on the relative importance of the interaction curve features on the different
components of the solvation phenomenon are postponed to a following section.}


\heading{c. Excitation spectra of Mg in He$_n$}

As mentioned in the introduction, the experimental spectra for the $^1$P$\leftarrow ^1$S excitation of Mg
dispersed in superfluid bulk He \cite{morika99} and attached to He nanodroplets \cite{reho00} are available.
Their features, as well as the comparison with excitation spectra of the free Mg atom, strongly suggest
that Mg is soluble in superfluid helium.
In this Section, we discuss the simulation results obtained employing the various ground state interaction 
curves for MgHe as a function of the number of helium atoms, and compare the experimental spectra with the 
simulated ones for the largest clusters available.

Figures 10a, 10b and 10c show the spectra computed with the MP4, CCSD(T) and CCSDT PES, respectively. In 
all cases, it is clearly evident an increasing blue-shift and FWHM of the absorption peak
upon increase of the number of He atoms in the clusters.
This finding is in line with what previously found for helium clusters doped with alkali metals \cite{yamashita}
and with atomic silver \cite{maxag}. However, differently from the AgHe$_n$ case, no red-shift is present for the
small clusters. This result, in conjunction with the relative shape of the ground and excited MgHe potentials,
suggests that the direct formation of exciplexes \cite{morika99} during the vertical excitation process is quite 
unlikely. We also notice that the magnitude of the blue-shift is larger for Mg than in the case of alkali-doped 
helium clusters \cite{yamashita}, indicating that the excited Mg atom finds itself in a more repulsive
environment, and the characteristic long tail of floating dopant is absent even in the CCSD(T) results.

Despite the similar trend observed upon increasing the number of He atoms, the three sets of spectra
present several interesting differences for a given cluster size: both the blue-shift and the FWHM 
increase in the sequence CCSD(T) $<$ CCSDT $<$ MP4. Furthermore, the CCSD(T) and CCSDT sets 
of results show a spectral shape different from the absorption curve obtained with the MP4 for some of the 
medium-size droplets, the MP4 spectra presenting a shoulder 
in the absorption curve at large wave-numbers already for MgHe$_{30}$. This feature is evident 
in the CCSDT case only starting from MgHe$_{40}$, and is never present in the CCSD(T) results.

Given the previous discussion on the dopant location obtained from the simulations, it is straightforward to
suggest that this shoulder is due to the onset of a more compact and complete first solvation
shell of Mg at $n=30$ for MP4 and at $n=40$ for CCSDT. A similar explanation was previously proposed 
to rationalize the increase in blue-shifts and in FWHM in AgHe$_n$ \cite{maxag}. 
More precisely, we recall that some of the DMC simulations generated
an anisotropic He environment around the dopant. In these systems, Mg resides
preferentially in a deep dimple on the cluster surface, and the local environment allows the
excited $^1$P state to orient itself so that the 3p orbital containing the excited electron is directed outside 
the cluster. As a consequence of this freedom, the repulsive interaction between the $^1$P Mg state and the 
local He environment is decreased (we recall that the $\Sigma$ excited state is highly repulsive), and so is 
the shift of the component of the spectral line due to the $\Sigma$ state \cite{shoulder}.
In turn, this delays the onset of the shoulder at large wave-numbers until the complete solvation has occurred.

The location of the maximum in the spectra can be used to better quantify the physical picture of 
the vertically excited Mg and to compute its binding energy to the cluster using the formula
$E_{bind}^{exc}=E_{bind}^{gs}-(h\nu_{cluster}-h\nu_{vac})$. Extracting the position of the maxima for the
simulations on MgHe$_{50}$ and using the data in Table \ref{tab3}, one obtains 
$E_{bind}^{exc}(\mathrm{MP4})\simeq-518$ cm$^{-1}$,
$E_{bind}^{exc}(\mathrm{CCSDT})\simeq-589$ cm$^{-1}$ and $E_{bind}^{exc}(\mathrm{CCSD(T)})\simeq-235$ cm$^{-1}$,
clearly indicating the instability of the excited Mg atom attached to
the clusters. In turn, these results suggest that the break up of the excited system into the excited Mg
atom and the ground state He cluster is a likely dissociation channel, and that the excited Mg could 
leave the droplet converting some of the excess electronic energy into kinetic energy. On the other hand,
another possible outcome following the vertical excitation is represented by a rearrangement of the He atoms 
surrounding the excited dopant; these atoms would adjust their distribution to reflect the new shape of the 
interaction potential and to produce a more anisotropic, but also more strongly bound,
environment around the Mg. During this process, some of the excess energy contained in the system after the excitation
must be somehow dissipated, and it is likely that this
would produce a partial break up of the He moiety. 
Unfortunately, it is difficult to predict the likelihood of
this process using exclusively our DMC results, and a dynamical simulation of the relaxation process
of the He cluster appears mandatory to clarify this detail. 

Comparing the experimental spectra reported in Refs. \cite{morika99} and \cite{reho00} with the simulation results 
obtained for our largest clusters (collected together in Fig. 10c), several interesting observations can be made. 
First, all the results show a good agreement concerning the location of the two experimental 
spectra, once again indicating the good accuracy of the DIM model to approximate the interaction energy for 
the dopant excited state. However, the shape of the MP4 and CCSDT spectra for MgHe$_{50}$ are much closer 
to the experimental bulk-He spectrum \cite{morika99} than to the one obtained in clusters
\cite{reho00}; the latter presents a convex shape in the region 35800-36500 cm$^{-1}$ where
the bulk spectrum shows a concave behavior (henceforth, a "shoulder").
It is also interesting to note that none of our simulation
results present an absorption shape similar to the cluster spectra 
in the region 35300-35800 cm$^{-1}$, where the experimental
results were fitted using two Gaussians; this feature was explained invoking the quadrupolar
deformation of the Mg cavity similarly to what was previously found for heavy alkali and Ag.
Since it was previously shown that coupling the DIM method with DMC or PIMC simulations \cite{maxag} 
accurately describes the features due to the anisotropy in the local dopant environment, at this stage it is
difficult for us to propose an explanation for the lack of structure found in the range 35300-35800 cm$^{-1}$.
The possibility to invoke the presence of a vibronic coupling between the He atoms and the 
Mg electrons during the excitation process seems to be ruled out by the lack of 
similar features in the bulk-He spectrum.
As an alternative, we can conceive that the different shape of the excitation spectra in
clusters and bulk may be due to the two different detection mechanisms used; this implies that 
the shape of the fluorescent emission is strongly dependent on the wavelength used to 
excite the chromophore, though.

On the computational side, we cannot rule out {\it a priori} the possibility that the 
absence of structure in the excitation spectrum may be due to
size effects and that larger clusters may indeed have this very feature.  However, both the 
very similar spectral shape obtained for MgHe$_{30}$, MgHe$_{40}$ and MgHe$_{50}$ using MP4,
and the almost saturated binding energy, strongly suggest that the local Mg environment 
-and hence the excitation spectrum- is converged with respect to the size of the system.

\heading{d. Analysis of the solvation parameters}

The DMC simulation results presented in the previous Sections have highlighted a dependency of the solubility 
and the spectral properties of Mg in He$_n$ clusters on the shape of the interaction curve for $6 \leq n \leq 50$.
However, a deeper quantitative understanding of the underlying causes producing
such distinct behaviours is still missing, in spite of the accurate numerical results.
In particular, we would like to understand which specific feature, if any, in the 
three potentials is primarily responsible for the change in solubility
highlighted by the Mg-GC distance and $\cos(\mathrm{HeMgHe})$ distribution functions.

To shed some light on the peculiarities of this phenomenon, we turned to an exact formulation of Density 
Functional Theory (DFT) as a way to extract energetic quantities that are
expected to play an important role in defining the impurity solvation.
To this end, we extended the exact DFT formalism 
for homogeneous systems proposed by Levy, Perdue and Sahni \cite{levy84} to describe
an impurity attached to a homogeneous cluster.
More specifically, we derived an effective one-particle Schr\"odinger equation for
the Mg-GC radial probability density and the Mg binding energy $\mu$ to the droplet containing
the effective potential 
\begin{equation}
V_{eff}(\mathbf{R}_{\mathrm{gcMg}})\simeq
V_{eff}^{\mathrm{HeHe}}(\mathbf{R}_{\mathrm{gcMg}}) +
V_{eff}^{\mathrm{MgHe}}(\mathbf{R}_{\mathrm{gcMg}})
\label{mgeq7}
\end{equation}
\noindent
where
\begin{equation}
V_{eff}^{\mathrm{HeHe}}(\mathbf{R}_{\mathrm{gcMg}})=
\frac{\langle \Phi | V_{\mathrm{HeHe}}(\mathbf{y}_{\mathrm{He}}) |\Phi \rangle_{n-1}}
{\langle \Phi| \Phi \rangle_{n-1}}
\label{mgeq7_bis}
\end{equation}
\noindent
and
\begin{equation}
V_{eff}^{\mathrm{MgHe}}(\mathbf{R}_{\mathrm{gcMg}})=
\frac{\langle \Phi | V_{\mathrm{MgHe}}(\mathbf{y}_{\mathrm{He}},\mathbf{R}_{\mathrm{gcMg}}) |\Phi \rangle_{n-1}}
{\langle \Phi| \Phi \rangle_{n-1}}
\label{mgeq7_tris}
\end{equation}
Here, $\mathbf{y}_{\mathrm{He}}$ indicates the 3$(n-1)$ dimensional vector describing the He atoms position
with respect to the centre of the He moiety and $\langle ... \rangle_{n-1}$ represents the average over
$\mathbf{y}_{\mathrm{He}}$.
For clarity of discussion, we present the complete derivation of this equation in Appendix A
and only mention that the two terms in the righthand side of Eq. \ref{mgeq7} represent the
average interaction of the He atoms among themselves and with Mg,
as a function of the position of the dopant in the cluster. Both quantities can be exactly or accurately computed
with DMC simulations.

Figure 11 shows the radial dependency of the angular average of $V_{eff}(\mathbf{R}_{\mathrm{gcMg}})$,
$V_{eff}^{\mathrm{HeHe}}(\mathbf{R}_{\mathrm{gcMg}})$ and
$V_{eff}^{\mathrm{MgHe}}(\mathbf{R}_{\mathrm{gcMg}})$
obtained during the simulation of MgHe$_{30}$ with the MP4 and CCSD(T) pair potentials. Basing on the results 
shown in Figure 7, this cluster and these two PES were chosen as representative cases of the possible different 
behaviours. 

The computed $V_{eff}^{\mathrm{MgHe}}$ values
decrease for $\| \mathbf{R}_{\mathrm{gcMg}}\| \rightarrow 0$, and this
is due to the increased number of He atoms surrounding the dopant  \cite{lehmann99}; in turn, this 
indicates the increased stabilization of the impurity when brought inside the cluster.
$V_{eff}^{\mathrm{MgHe}}$ also shows a parabolic shape
when $\| \mathbf{R}_{\mathrm{gcMg}}\| \simeq 0$, strongly suggesting that the analysis 
carried out by Lehmann \cite{lehmann99} is valid for small clusters.
Two differences are however present for the two PES, namely a global shift in value reflecting
the difference in the well depth between MP4 and CCSD(T),
and a larger curvature for the MP4 results; the latter indicates a tighter
binding of the impurity to the cluster centre in the MP4 case.

At variance with $V_{eff}^{\mathrm{MgHe}}$,
$V_{eff}^{\mathrm{HeHe}}$
increases in value when the impurity is closer to the geometrical centre of the 
He moiety, a finding that can be easily explained by recalling that the He atoms lose 
stabilizing interactions due to the formation of a cavity inside the cluster and to 
the increase of the cluster surface area. Similarly to the discussion given by
Lehmann \cite{lehmann00} on the so called "buoyancy correction", we also found that
the He atoms lose more 
stabilizing interactions when the impurity sits close to the cluster centre than when it is 
just below the surface of the droplet. In this respect, the CCSD(T) results show
a steeper increase in $V_{eff}^{\mathrm{HeHe}}$ than the MP4 ones,
the most likely explanation for this feature being the different effective radius for Mg 
produced by the two potentials. Defining this radius as the distance at which a given PES has a value of zero,
we found that the CCSD(T) PES is larger by roughly 0.2 bohr than the MP4 one. 

Due to the different features of $V_{eff}^{\mathrm{HeHe}}$ and $V_{eff}^{\mathrm{MgHe}}$ 
as a function of the interaction curve, $V_{eff}$ shows two qualitatively different shapes:
the CCSD(T) potential presents an overall repulsive interaction from the cluster surface inward
due to the large values of $V_{eff}^{\mathrm{HeHe}}$, while
the MP4 potential shows an almost complete cancellation between the two components and produces a shallow 
attractive interaction between the dopant and the centre of the He moiety. In the latter case, it should be
stressed that both the larger interaction between Mg and He and the smaller size of Mg play a relevant 
role in defining the shape of the total effective potential.
$V_{eff}$ was also computed using the MP4 curve
for MgHe$_{12}$, MgHe$_{20}$, MgHe$_{30}$ and MgHe$_{40}$ and gave
a qualitatively correct prediction of the solvation phenomenon. Indeed, whereas
it presents a deep minimum away from the droplet centre in the small clusters,
a shallow minimum close the centre of the He moiety is produced upon increasing the system size. 



\heading{V. CONCLUSIONS}

In this work we presented a computational study of the ground state properties and excitation spectrum 
of Mg attached to He$_n$ clusters ($n=2-50$). This study was carried out using the exact DMC method in conjunction 
with accurate model potentials written as a sum over pair interactions, a commonly used and fairly accurate 
approximation for weakly interacting systems. 
For the He-He interaction potential we used the well known and widely employed TTY model \cite{tang95}, whereas
large basis set {\it ab initio} MP4 and CCSD(T) energies were computed and interpolated using quintic splines
to produce two accurate potentials for the Mg-He pair. Additionally, a third Mg-He model potential was obtained by 
interpolating the CCSDT interaction energies computed by Hinde \cite{hinde03}. The well depth and equilibrium 
distance of the ground state PES are -5.70 cm$^{-1}$ and 5.03 {\AA} for MP4, -4.74 cm$^{-1}$ and 5.11 {\AA} for
CCSD(T), and -5.01 cm$^{-1}$ and 5.07 {\AA} for CCSDT computations.

We computed total, evaporation and Mg binding energies as a function of $n$ using the three interaction curves; 
the comparison between the results for MgHe$_{40}$ and MgHe$_{50}$ indicates an almost converged He environment around 
the atomic dopant for $n=50$. In spite of the similar shape of the interaction curves (Fig. 1), the results of the 
DMC simulations revealed markedly different solubilities of Mg in He$_n$ as a function of both the Mg-He pair 
interaction and $n$: the MP4 PES suggests that 
Mg is located in the centre of He$_n$ independently of $n$ ; CCSD(T) always predicts a surface location for the dopant;
CCSDT, instead, presents a change in Mg behavior around $n=25$, predicting Mg to be soluble in the larger clusters.
These results highlight both a marked sensitivity of the solvation properties with respect to the global shape of the 
PES and the ability of DMC to discriminate qualitatively different behaviors generated by subtle differences in
the interaction potentials. To gain a deeper understanding of the solvation process in these species, we used a 
modified version of the DFT approach proposed in Ref. \cite{levy84} and found
that both the well depth and the position of the repulsive wall influence the solubility of Mg in He$_n$.

Our simulation results bear relevance also with respect to the solvation model proposed by Ancilotto {\it et al.}
\cite{anc}. This model predicts solvation whenever the dimensionless parameter $\lambda$ exceeds 1.9, whereas our 
results present a quite different
scenario: the simulations using the CCSD(T) PES strongly suggest that solvation is not automatically assured
in spite of the large $\lambda$ ($\lambda_{\mathrm{}CCSD(T)} = 2.66$). Furthermore, the dopant 
solubility may depend on the size of the cluster (as highlighted by the CCSDT simulations) and be sensitive to
subtle differences between interaction potentials (for instance, compare the CCSD(T) and CCSDT case).
In turn, this findings stress that highly accurate pair interactions are needed to correctly predict the solubility
of neutral dopants with isotropic pair potential.

We found a dependency on the shape of the PES also for the simulated excitation spectra, the solvated Mg (MP4 and CCSDT
cases) showing a broader and more blue-shifted absorption band than the floating dopant (CCSD(T)). Differently 
from the results obtained for AgHe$_{n}$ \cite{maxag}, no red-shifted bands were found for the small MgHe$_n$ 
clusters, suggesting that the formation of exciplexes would require
a re-organization of the He environment around Mg following 
the vertical excitation. Indeed, this outcome is considered quite likely by Reho {\it et al.} \cite{reho00} on 
the basis of their time-resolved results, and it would be interesting to study the dynamical changes involved
during this process.

Comparing the simulated and experimental cluster spectra we found a good agreement between the position of the 
excitation bands, once more validating the usage of the DIM method to approximate the excited PES. 
In spite of this 
agreement, there is a marked difference in shape between the experiments and the results obtained by theory 
for MgHe$_{50}$: whereas the simulated spectra always present a smooth band without peculiar features, the 
experimental spectrum shows a two-peaks structure interpreted as the outcome of a quadrupolar deformation of the
He cage around the solvated Mg. Bearing in mind that DMC was previously found to describe correctly this kind of 
deformations \cite{maxag}, this discrepancy calls for further theoretical and experimental investigations.

\heading{ACKNOWLEDGMENTS}
MM acknowledges Dario Bressanini for independently testing some of the simulation results and for providing
a more precise DMC energy for He$_{40}$, and the EPSRC for an Advanced Research Fellowship (GR/R77803/01).

\clearpage
\heading{Appendix A. Effective Schr\"odinger equation for a single atomic impurity in He clusters}

The present derivation of the effective Schr\"odinger equation for an atomic dopant (e.g. Mg) 
attached to He$_n$ is a modification of the derivation presented by Levy {\it et al.} \cite{levy84} 
for a homogeneous many-particle system.

As first step, the Schr\"odinger equation for a generic MgHe$_n$ cluster with
the Hamiltonian operator shown in Eq. \ref{mgeq2} must be rewritten in Jacobi coordinates 
$\mathbf{y}_{j-1}=\mathbf{r}_j-\frac{\sum_i^{j-1}m_i \mathbf{r}_i}{\sum_i^{j-1}m_i}$
where $\mathbf{r}_i$ ($i \leq n$) is the position of the $i$-th He atom, $m_i$ its mass,
and $\mathbf{r}_{n+1}$ and $m_{n+1}$ are the position and mass of Mg, respectively.
Using these definition, $\mathbf{y}_{n}$ and $\mu_{n}$ represent the distance and reduced mass of the impurity 
with the geometrical centre of the He moiety, while $\mathbf{y}_{n+1}=\mathbf{R}_{\mathrm{COM}}$ 
is the position of the MgHe$_n$ centre of mass in the laboratory frame.
It is also worth pointing out that the set of coordinates 
$\mathbf{y}_{\mathrm{He}}=(\mathbf{y}_{1},...,\mathbf{y}_{n-1})$ describes the relative
positions of the He atoms within the helium moiety.

After the rotation in Jacobi's coordinates, the Hamiltonian for the cluster can be written as

\begin{equation}
\mathcal{H} = -\frac{1}{2} \sum _{j=1} ^{n}
\frac{\nabla _{j} ^{2}}{\mu_{j}}
 -\frac{\nabla _{\mathrm{COM}} ^{2}}{2M_{\mathrm{COM}}}
+ V_{\mathrm{HeHe}}(\mathbf{y}_1,...,\mathbf{y}_{n-1})
+ V_{\mathrm{MgHe}}(\mathbf{y}_1,...,\mathbf{y}_{n})
\label{appen1}
\end{equation}
\noindent
where $\mu_{j}=\frac{m_{j+1}M_{j}}{M_{j+1}}$,
$M_{j}=\sum_{i=1}^{j} m_i$, and
$M_{\mathrm{COM}}=M_{n+1}$ is the total mass of the system, 

Neglecting the term $-\frac{\nabla _{\mathrm{COM}} ^{2}}{2M_{\mathrm{COM}}}$ from
Eq. \ref{appen1} allows to eliminate the contribution of the centre of mass kinetic energy to
the system total energy, so that
the formal ground state solution of this equation 
$\Psi(\mathbf{Y})$ ($\mathbf{Y}=(\mathbf{y}_1,...,\mathbf{y}_{n})$) will describe only the
internal motion of the system.

To proceed further, we partition the total Hamiltonian into the two operators
\begin{equation}
\mathcal{H}_0(\mathbf{y}_{\mathrm{He}},\mathbf{y}_{n}) = -\frac{1}{2} \sum _{j=1} ^{n-1}
\frac{\nabla _{j} ^{2}}{\mu_{j}} 
+ V_{\mathrm{HeHe}}(\mathbf{y}_1,...,\mathbf{y}_{n-1})
+ V_{\mathrm{MgHe}}(\mathbf{y}_1,...,\mathbf{y}_{n})
\label{appen2}
\end{equation}
\noindent
and $-\frac{\nabla _{n} ^{2}}{2\mu_{n}}$, the latter representing the kinetic energy of the
relative motion of the impurity with respect to the centre of mass of the He moiety.
The exact ground state wave function can also be arbitrarily written as

\begin{equation}
\Psi(\mathbf{Y})=\rho^{1/2}(\mathbf{y}_{n})\Phi(\mathbf{Y})
\label{appen3}
\end{equation}
\noindent
where $\rho(\mathbf{y}_{n})$ is the probability density of finding the impurity at position
$\mathbf{y}_{n}$ with respect to the centre of the He cluster.

A series of algebraic steps are now necessary. First, let us
indicate by $E^n_0(\mathrm{Mg})$ and $E^{n}_0(\mathrm{pure})$ the ground state energy for 
MgHe$_n$ and He$_{n}$, respectively.
Inserting Eq. \ref{appen3} into the Schr\"odinger equation obtained eliminating the
centre of mass kinetic energy from the Hamiltonian in Eq. \ref{appen1},
subtracting $E^{n}_0(\mathrm{pure})\rho^{1/2}(\mathbf{y}_{n})\Phi(\mathbf{Y})$ 
from both sides, and integrating both sides of
the Schr\"odinger equation over $d\mathbf{y}_{1}...d\mathbf{y}_{n-1}\Phi(\mathbf{Y})$ leads to

\begin{eqnarray}
-\langle \Phi| \Phi \rangle_{n-1}\frac{1}{2\mu_{n}}\nabla _{n}^2\rho^{1/2}(\mathbf{y}_{n})+
\rho^{1/2}(\mathbf{y}_{n})\langle \Phi | \mathcal{H}_0(\mathbf{y}_{\mathrm{He}},\mathbf{y}_{n}) |\Phi \rangle_{n-1} \\ \nonumber
-\frac{1}{\mu_{n}}\nabla _{n}\rho^{1/2}(\mathbf{y}_{n}) \langle \Phi(\mathbf{Y})\nabla _{n} 
\Phi(\mathbf{Y})\rangle_{n-1} 
-\frac{1}{2\mu_{n}}\rho^{1/2}(\mathbf{y}_{n})\langle \Phi(\mathbf{Y})\nabla _{n}^2 \Phi(\mathbf{Y})\rangle_{n-1} \\ \nonumber
-E^{n}_0(\mathrm{pure})\rho^{1/2}(\mathbf{y}_{n})\langle \Phi| \Phi \rangle_{n-1}=
(E^n_0(\mathrm{Mg})-E^{n}_0(\mathrm{pure}))\rho^{1/2}(\mathbf{y}_{n})\langle \Phi| \Phi \rangle_{n-1}
\label{appen4}
\end{eqnarray}
\noindent
where $\langle ... \rangle_{n-1}$ indicates the integration over $d\mathbf{y}_{1}...d\mathbf{y}_{n-1}$.

Recognizing that $E^n_0(\mathrm{Mg})-E^{n}_0(\mathrm{pure})=\mu_{\mathrm{Mg}}$ is the chemical potential 
of the Mg impurity attached to the He$_n$ cluster,
that $\langle \Phi(\mathbf{Y})\nabla _{n} \Phi(\mathbf{Y})\rangle_{n-1}$ is identically zero,
and dividing both sides of Eq. \ref{appen4} by $\langle \Phi| \Phi \rangle_{n-1}$, the previous equation can be written as

\begin{eqnarray}
-\frac{1}{2\mu_{n}}\nabla _{n}^2\rho^{1/2}(\mathbf{y}_{n})+
\rho^{1/2}(\mathbf{y}_{n})\frac{\langle \Phi | \mathcal{H}_0(\mathbf{y}_{\mathrm{He}},\mathbf{y}_{n})-E^{n}_0(\mathrm{pure}) |\Phi \rangle_{n-1}
-\frac{1}{2\mu_{n}}\langle \Phi(\mathbf{Y})\nabla _{n}^2 \Phi(\mathbf{Y})\rangle_{n-1}
}{\langle \Phi| \Phi \rangle_{n-1}}= \nonumber \\
-\frac{1}{2\mu_{n}}\nabla _{n}^2\rho^{1/2}(\mathbf{y}_{n})+ V_{eff}(\mathbf{y}_{n})\rho^{1/2}(\mathbf{y}_{n})
=\mu_{\mathrm{Mg}}\rho^{1/2}(\mathbf{y}_{n})
\label{appen5}
\end{eqnarray}
where it is shown that the exact density and chemical potential for the impurity can be obtained
by solving an effective Schr\"odinger equation for the impurity alone. However, the 
effective potential $V_{eff}(\mathbf{y}_{n})$ is not know in advance, and must be obtained, in principle, 
by solving the many-body problem and computing the relevant quantities
over the exact wave function $\Psi(\mathbf{Y})$. Besides, it should be expected to be highly system-dependent,
so that a general form valid for a family of systems may be difficult to obtain.

From Eqs. \ref{appen5} and \ref{appen2}, we notice that $V_{eff}$ contains
components deriving from the average values of the interaction potentials and of the 
kinetic energy operators, the latter being zero or constant 
if and only if the exact wave function can be factorized exactly into $\Phi(\mathbf{y}_{\mathrm{He}})$ 
and $\rho^{1/2}(\mathbf{y}_{n})$. Such an instance obviously implies a complete decorrelation between the
He atoms and dopant motion, a situation that cannot be encountered in practice because
the impurity always modifies both the value and the curvature of $\Phi$ as a function
of its position. However, neglecting this contribution may represent an accurate approximation in the case
of floating impurities spending most of the time away from the region of high He density in the cluster.

Due to the presence of these kinetic energy terms, the effective potential introduced in Eq. \ref{appen5} 
is different from the one previously defined by Lehmann \cite{lehmann99,lehmann00} to describe the energetic
effects of displacing an impurity from the droplet centre. In the formulation presented in 
Ref. \cite{lehmann99}, the change in curvature (hence in the kinetic energy) of the He wave function 
due to the location of the impurity was disregarded. Moreover, our definition of $V_{eff}$ takes naturally 
into account the local increase of He density around the impurity due to the interaction potential, a 
feature neglected in the simplified discussion made by Lehmann, and that we expect 
to play a fundamental role in defining the solubility of a neutral dopant.

Despite our ignorance about its form, the effective potential defined in Eq. \ref{appen5} may
still be an useful tool to provide insight in the dopant behavior as it is easy to
show that $V_{eff}$ can be computed during a QMC simulation

\clearpage

\bibliography{bsample}

\begin{thebibliography}{10}
\bibitem{surface_volume}
Assuming the liquid He droplet to behave as an uncompressible fluid, it can be
  shown that the increase in the droplet surface $S$ due to the solvation of a
  rigid impurity or radius $r$ is given by $\Delta S(V,r)=4\pi
  V^{2/3}{[3/(4\pi)+r^3/V]^{2/3}-[3/(4\pi)]^{2/3}}$, where $V$ is the original
  volume of the droplet. For large $r^3/V$ ratios, $\Delta S(V,r)$ can be shown
  to decay as $V^{-1/3}$ indicating a smaller surface increase for larger $V$.
\bibitem{shoulder}
We tested this idea by decomposing the spectrum in its three components due to
  the three eigenvalues obtained from the DIM matrix. Invariably, we found that
  the one with the largest blue-shift was primarly due to the $\Sigma$ state
  interactions and that it was responsible for the right shoulder on the
  spectrum peak.
\end{thebibliography}

\clearpage

\begin{table}
\begin{center}
\begin{tabular}{lcccc}  \hline \hline
Mg & He & Bond functions & $D_e$ (cm$^{-1}$) & r$_e$ (A)\\ \hline
6-311+G3df& d-aug-cc-pVQZ& -& -5.21& 5.09\\
6-311+G3df& d-aug-cc-pV5Z &- &-5.36 &5.07\\
A.N.O. triple zeta \cite{widmark91}&d-aug-cc-pVQZ &- &-5.34 &5.10\\
A.N.O. triple zeta &d-aug-cc-pV5Z &- &-5.44 &5.08\\
6-311+G3df &d-aug-cc-pVTZ &332 &-5.63 &5.07\\
6-311+G3df &d-aug-cc-pVQZ &332 &-5.74 &5.07\\
6-311+G3df& d-aug-cc-pV5Z &332 &-5.76 &5.06\\
6-311+G3df &d-aug-cc-pVQZ &33221 &-5.77 &5.06\\
6-311+G3df &d-aug-cc-pV5Z &33221 &-5.78 &5.06\\
A.N.O. triple zeta &d-aug-cc-pVQZ &332 &-5.64 &5.08\\
A.N.O. triple zeta &d-aug-cc-pVQZ &33221 &-5.67 &5.08\\
Archibong \cite{archi91}&d-aug-cc-pVQZ &332 &-5.79 &5.06\\
Archibong &d-aug-cc-pV5Z &332 &-5.82 &5.06\\
Archibong &d-aug-cc-pVQZ &33221 &-5.84 &5.06\\
\hline \hline
\end{tabular}
\caption{Effect of the basis set choice on the equilibrium properties of the
MgHe dimer. The different basis sets are introduced in the text.
The estimate of the equilibrium parameters are obtained by interpolation with a quadratic polynomial
of the energy data at 4.75, 5.00, and 5.25 {\AA}.}
\label{tab1}
\end{center}
\end{table}

\clearpage

\begin{table}
\begin{center}
\begin{tabular}{lcccccc}  \hline \hline
$n$ & $E_{DMC}^{\mathrm{MP4}}$ & $\Delta_{DMC}^{\mathrm{MP4}}$ & $E_{DMC}^{\mathrm{CCSD(T)}}$ & $\Delta_{DMC}^{\mathrm{CCSD(T)}}$ &$E_{DMC}^{\mathrm{CCSDT}}$ & $\Delta_{DMC}^{\mathrm{CCSDT}}$\\ \hline
 2  & -2.5366(6) & -1.2683(3) &            &           &     &     \\
 4  & -5.660(2)  & -1.5618(8) & -4.127(4)  & -1.0317(4)&     &     \\
 6  & -9.2445(9) & -1.7921(9) & -6.9881(9) & -1.431(1) &     &     \\
 8  & -13.172(5) & -1.964(2)  & -10.244(5) & -1.628(3) &     &     \\
12  & -21.795(3) & -2.156(1)  & -17.619(6) & -1.843(2) & -18.406(4) &     \\
15  & -28.84(1)  & -2.346(3)  & -23.717(3) & -2.032(2) & -24.648(7) & -2.081(2)  \\
18  &  -36.33(1) & -2.498(5)  & -30.26(2)  & -2.182(7) & -31.26(1)  & -2.204(3) \\
20  &  -41.54(2) & -2.61(1)   & -34.76(2)  & -2.25(1)  & -35.894(7) & -2.317(5)   \\
25  &  -55.031(8)& -2.698(4)  & -46.73(2)  & -2.394(5) & -47.984(9) & -2.418(2)   \\
30  & -68.75(3)  & -2.744(6)  & -59.37(3)  & -2.526(7) & -60.75(1)  & -2.553(2)    \\
40  & -97.13(5)  & -2.838(6)  & -85.88(3)  & -2.651(4) & -87.93(2)  & -2.718(2)    \\
50  & -126.70(9) & -2.96(1)   & -114.27(5) & -2.839(5) & -116.84(3) & -2.891(4) \\
\hline \hline
\end{tabular}
\caption{DMC energy $E_{DMC}(n)$ and $\Delta(n)= [E(n)-E(m)]/(n-m)$ as a function of
the number of He atoms in the cluster for the three interaction potentials. 
For a given $n$, MgHe$_m$ is the largest cluster available with $n>m$. Energetic
quantities are in cm$^{-1}$.
}
\label{tab2}
\end{center}
\end{table}

\clearpage
\begin{table}
\begin{center}
\begin{tabular}{lccc}  \hline \hline
$n$ & $E_{DMC}^{\mathrm{MP4}}$ & $E_{DMC}^{\mathrm{CCSD(T)}}$ & $E_{DMC}^{\mathrm{CCSDT}}$ \\ \hline
 2  & 2.5357(6) &           &           \\
 4  & 5.271(2)  & 3.738(4)  &           \\
 6  & 7.6368(9) & 5.3804(9) &           \\
 8  & 9.604(5)  & 6.676(5)  &           \\
12  & 13.049(7) & 8.873(6)  & 9.660(7)  \\
20  &  18.50(2) & 11.72(2)  & 12.85(1)  \\
25  &  24.4(3)  & 16.1(3)   & 17.4(3)   \\
30  & 26.69(3)  & 17.31(3)  & 18.69(3)  \\
40  & 26.85(5)  & 15.60(3)   & 17.65(3)  \\
\hline \hline
\end{tabular}
\caption{Binding energy of Mg to He$_n$ as a function of the number $n$
of He atoms in the cluster for the three interaction potentials.
Quantities are in cm$^{-1}$.
}
\label{tab3}
\end{center}
\end{table}

\clearpage
\noindent
{\bf Figure captions:}\\
\noindent
Figure 1: 
Ground state interaction energy curves for Mg-He and He-He. The CCSDT potential is 
taken from Ref. \cite{hinde03}. Energies in cm$^{-1}$ and
distances in atomic units. \\ \noindent
Figure 2: 
$^1\Pi$ interaction energy curve between $^1$P Mg and He as a function of the method
(FCI or MRCI) and of the basis set used. Energies in cm$^{-1}$ and
distances in atomic units. \\ \noindent
Figure 3: 
$^1\Sigma$ interaction energy curve between $^1$P Mg and He as a function of the method
(FCI or MRCI) and of the basis set used. Energies in cm$^{-1}$ and
distances in units. \\ \noindent
Figure 4: 
Two-body (empty symbols) and three-body (filled symbols) effects computed at the MP4 level
for MgHe$_2$ complexes in isosceles triangular geometry, as a 
function of the MgHe distance and the HeMgHe angle. Energies in cm$^{-1}$, distances in {\AA} and
angles in degrees.\\  \noindent
Figure 5:
Evaporation energy $\Delta(n)$ as a function of the interaction potential and
of the number of helium atoms for MgHe$_n$. The $\Delta(n)$ values for pure
He$_n$ are also included. Energies in cm$^{-1}$. \\ \noindent
Figure 6:
Average distance (bohr) between Mg and the geometrical centre (GC) of the He moiety 
as a function of the number of He atoms in the cluster and of the interaction 
potential.\\ \noindent
Figure 7:
Mg radial probability density function with respect the geometrical centre (GC) of the He moiety
as a function of the number of He atoms in the cluster and of the interaction
potential: a) MP4 b) CCSD(T) c) CCSDT. Distances in bohr.\\ \noindent
Figure 8:
Probability density function for the cosine of the He-Mg-He angle
as a function of the number of He atoms in the cluster and of the interaction
potential: a) MP4 b) CCSD(T) c) CCSDT. Each curve has been shifted upward by 0.1 with 
respect to the previous one to facilitate the comparison.\\ \noindent
Figure 9:
Snapshots of the walker distributions for MgHe$_{50}$ extracted during the DMC simulations using
a) the MP4, b) the CCSD(T) and c) the CCSDT pair interactions. Panel a) evidences the complete
solvation of the Mg atom inside the He cluster. Panel b) clearly corresponds to Mg resident in a deep dimple on
the surface, while c) indicates the typical asymmetric distribution of the He atoms around the Mg dopant
obtained with the CCSDT model.\\ \noindent
Figure 10:
Simulated excitation spectra of Mg attached to He$_n$ clusters. a) Results obtained using the MP4 potential for
$n=$ 12, 15, 20, 30, 40 and 50. b) Results obtained using the CCSD(T) potential for
$n=$ 18, 20, 25, 30, 40, and 50. c) Results obtained using the CCSDT potential for
$n=$ 25, 30, 40, and 50. The results obtained with MP4 and CCSD(T) for MgHe$_{50}$ are also reported in this panel
for a direct comparison.\\ \noindent
Figure 11:
Radial dependency of the angular average of the total effective potential $V_{eff}$ 
and of its components $V_{eff}^{\mathrm{HeHe}}$ and $V_{eff}^{\mathrm{MgHe}}$ for 
MgHe$_{30}$ using the MP4 and CCSD(T) potential curves. \\ \noindent

\clearpage

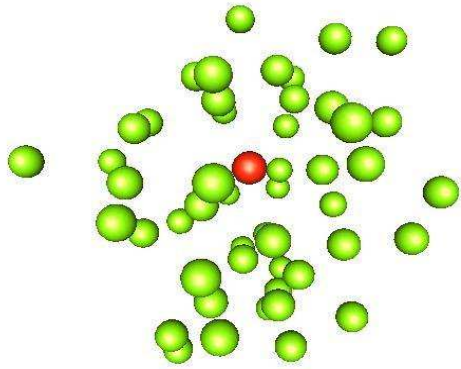
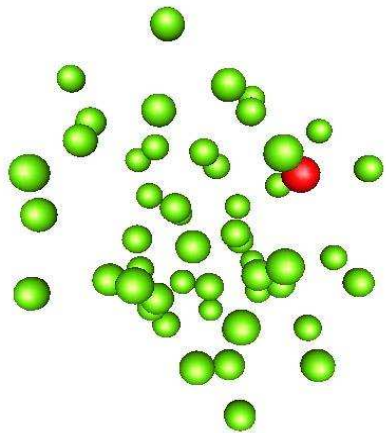
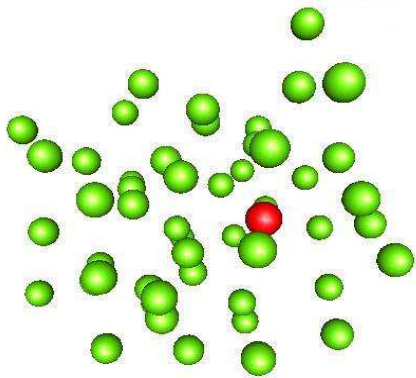
\end{document}